\documentclass[twocolumn,preprintnumbers,amsmath,amssymb]{revtex4-2}

\usepackage{graphicx}  
\usepackage{sidecap}
\begin{document}
%
\title{Resonant plasmonic terahertz detection  in  gated graphene $p-i-n$ field-effect structures enabled by the Zener-Klein tunneling nonlinearity
}
\author{V.~Ryzhii$^{1}$, T.~Otsuji$^1$, M.~Ryzhii$^{2}$,
V.~Mitin$^3$, and M.~S.~Shur$^4$}
\address{
$^1$Research Institute of Electrical Communication,~Tohoku University,~Sendai~ 980-8577, Japan\\
$^2$Department of Computer Science and Engineering, University of Aizu, Aizu-Wakamatsu 965-8580, Japan\\
$^3$Department of Electrical Engineering, University at Buffalo, SUNY, Buffalo, New York 14260 USA\\
$^4$Department of Electrical,~Computer,~and~Systems~Engineering, Rensselaer Polytechnic Institute,~Troy,~New York~12180,~USA
}

\begin{abstract} 
\normalsize
We show that resonant plasmonic detection dramatically increases the sensitivity of the 
terahertz  detectors based on a gated graphene $p-i-n$ (GPIN)  field-effect transistor (FET) structure.
In the proposed device, the gated $p$ and $n$ regions serve as the hole and electron reservoirs and the THz resonant plasma cavities. The current-voltage  ($I-V$) characteristics are strongly nonlinear due to the Zener-Klein interband tunneling in the reverse-biased i-region between the gates. The THz signal rectification by this region enables the THz detection. 
 The resonant excitation of the hole and electron   plasmonic oscillations
results in a substantial increase in the terahertz detector responsivity at the signal frequency close to the plasma frequency and its harmonics. 
Due to the transit-time effects, the GPIN-FET response at the higher plasmonic modes could be stronger than for the fundamental mode. Our estimates predict the detector responsivity up to a few of $10^5$~V/W at room temperature, muchlarger than for other electronic THz detectors, such as Schottky diodes, p-n-junctions, Si CMOS and III-V and III-N HEMTs. 
\end{abstract} 
\maketitle

\section{Introduction}
Short channel  field-effect transistors (FETs)
can serve as effective terahertz  detectors~\cite{1}. Such detectors could operate in a resonant regime when the detection is strongly enhanced by plasmonic resonances or in the rectification regime when the plasmonic oscillations are overdamped. In either case, the detector responsivity is proportional to the nonlinearity of the current-voltage ($I-V$) characteristics.
Different nonlinearity mechanisms enable the terahertz signal rectification in plasmonic FET detectors including the hydrodynamic nonlinearity and the barrier rectification in the Schottky junctions, $p-n$ junctions, or electrostatic barriers (see, for example,~\cite{2,3,4}).
The FET  detectors with the graphene layer (GL) channel can exhibit markedly enhanced performance~\cite{5,6,7,8,9,10,11,12} due to the unique electron and hole transport properties of GLs~\cite{13,14,15,16}, in particular, high electron (hole) mobility and directed velocity at elevated temperatures. The specific features of the GL band structure enable the Zener-Klein interband tunneling~\cite{17,18,19,20,21,22}  leading to a very strong nonlinearity that can be used for the rectification and detection of the terahertz signals.
In this paper, we evaluate the proposed THz detector based on a lateral graphene $p-i-n$ FET (GPIN-FET) detector structure with the gated p- and n-regions of the GL channel. 
This device combines the advantages of the strong $I-V$ nonlinearity and plasmonic resonant detection enhancement enabled
by high carrier mobility in GLs and transit-time effects. Such a combination of high nonlinearity and plasmonic effects leads to a remarkably high responsivity. Our estimates predict the detector responsivity up to a few  $10^5$~V/W,
markedly exceeding that of Schottky diodes, p-n junctions, Si CMOS, and III-V and III-N HEMTs.

\begin{figure}[b]\centering
\includegraphics[width=6.9cm]{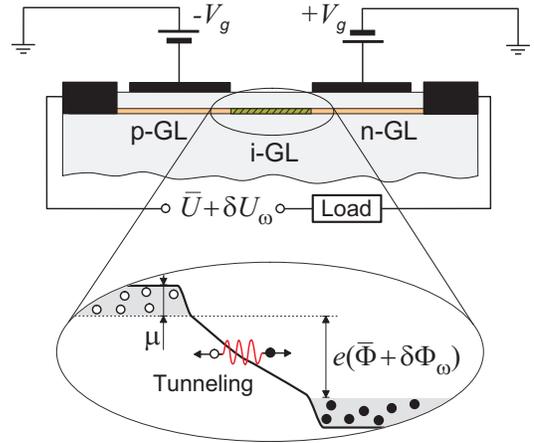}
\caption{ 
Schematic view of the cross-sections of the  GPIN-FET structure 
  with  gated electrostatically doped $p$ and $n$ regions in the  GPIN-FET channel. Inset shows
the   potential profile in the  channel.} 
\label{F1}
\end{figure}

\section{Model}

We consider the  GPIN-FET detector structure based on a GL  channel embedded in the dielectric [for examample hexagonal boron nitride  ($h$-BN)]. The channel is covered by two highly conducting gates. The gate voltages of different polarity, $\pm V_g$, are applied to  these gates. 
As a result,
the channel comprises the undoped $i$ regions between the gates (of the length $2l$) and the electrically doped p and n regions beneath the gates (of the length $L$).
The lengths of the $p$ and $n$ regions are close to the length of the  gates ($L$ is somewhat larger than the gate length 
due to the finite screening length or the gate fringe effect).
The channel is bounded by 
the source and drain contacts, between which the bias voltage ${\overline U}$ is applied.
Figure~1  shows the GPIN-FET structure with the electrically disconnected gates and   the potential profile in the GPIN-FET channel at the dc condition (inset)
when the bias voltage is applied between the side (source/drain) contacts.

In the present paper, 
 we consider the  GPIN-FET structures, 
where 
the Coulomb drag effect in the gated regions~\cite{23,24,25,26,27,28} is weak. This corresponds to GPIN-FET structures, in which
more  liberal and rather practical  conditions are fulfilled:
stronger scattering on  impurities, defects, and acoustic phonons (with the collision frequencies $ \nu \gtrsim  1$~ps$^{-1}$), and relatively long gated regions
($L \gtrsim 1~\mu$m). The $i$ region is assumed to be fairly short [($2l = 0.1 - 0.2~\mu$m),
so that the transport of the holes and electrons generated in this region due to the Zener-Klein tunneling is  ballistic~\cite{29,30,31,32,33}. Considering that the i region is depleted under the operational conditions, the conditions of the ballistic transport in this region with the above length
 can be realized even at room temperature~\cite{34} and, naturally, at lowered temperatures~\cite{35}.   
 The holes and electrons generated due to tunneling  are directed primarily along the electric field in the
$i-$region, i.e.,  in the $x$-direction (the in-plane direction along the GL channel) from the $p$ and $n $ region, and propagate ballistically  with the velocities, $v = \pm v_W$.

Apart from the dc bias voltage ${\overline U}$, an ac signal voltage $\delta U_{\omega} \exp(-i\omega t)$ 
is applied, where $\delta U_{\omega}$ and $\omega$ are the amplitude and frequency of the incoming terahertz signal. Thus, both the dc and ac voltage drops across  the GPIN-FET structure are equal 
to ${\overline V} = {\overline U} - {\overline U}^{load}$
and $\delta V_{\omega} = \delta U_{\omega} - \delta U_{\omega}^{load}$, where ${\overline U}^{load}$ and
 $\delta U_{\omega}^{load}$ are the  dc  and ac components of the load voltage. These components depend
 on the load impedance $Z^{load}$. 
The latter can be presented as $Z^{load} = [1/r^{load} -i\omega C^{load}]^{-1}$, where $r^{load}$
and $C^{load}$ are the load resistance and capacitance, respectively. We assume that $C^{load}$ is
sufficiently large, so that the ac voltage drop across it is negligibly small, therefore, 
$\delta U_{\omega}^{load} \simeq 0$ and $\delta V_{\omega} = \delta U_{\omega}$. The real part of the load impedance, i.e., its resistance, $r^{load}$ determines the dc output signal.

\section{Linear response}

When both the dc voltage  ${\overline V}$ and the periodic signal voltage $\delta V_{\omega}\exp(-i\omega t)$
drop
across the GPIN-FET intrinsic part [$V = {\overline V} +\delta V_{\omega}\exp(-i\omega t)$],
the carrier current density  in the $i$ region (and in  other sections of the channel) is equal to $J^i = {\overline J}^i + \Delta J^i$. Here ${\overline J}^i$ is the dc component. The  component $\Delta J^i$
comprises the linear ac component $\delta J_{\omega}^i$ proportional to $\delta V_{\omega}$
and the ac rectified component $\Delta J_{\omega}^i$. The latter is due to the nonlinear $I-V$
characteristics of the $i$ region at the conditions of the Zener-Klein tunneling.

The voltage across the devices  is distributed  between the $i$ region and the  gated regions.

\subsection*{Dynamic conductance of the $i$ region}

Considering the bladelike configuration of the conducting areas surrounding the $i$ region, the pertinent
spatial electric-field  distribution in this region~\cite{36,37,38,39}, and using the general formulas for the Zener-Klein tunneling probability in graphene~\cite{17,18,19}, 
one can arrive at the following expression for the 
dc current density per unit width in the direction along the gate edges~\cite{22}:

\begin{eqnarray}\label{eq1}
{\overline J} = a
\frac{ev_W}{\sqrt{2l}}\biggl(
\frac{e{\overline \Phi}}{\hbar\,v_W}\biggr)^{3/2},
\end{eqnarray} 
where $a$ is a numerical parameter. At moderate   ${\overline \Phi}$ when the carrier space
charge in the $i$ regions is weak,
$$ a =\displaystyle \frac{\Gamma(1/4)\Gamma(1/2)}{\Gamma(3/4)}\frac{1}{2\pi^{7/2}},
$$
where $\Gamma(z)$ is the Gamma function.
Equation~(1) corresponds to the i-region linear dynamic ac conductance (see Appendix A)

\begin{eqnarray}\label{eq2}
 \sigma_{\omega}^i =  \sigma^i F_{\omega}-i\omega\,c^i = \sigma^i (F_{\omega}-i\omega\,\tau^i).
\end{eqnarray}
Here, $\sigma^i =d{\overline J}^i/d{\overline \Phi}$ is the $i-$region DC differential conductance, 
$\tau^i = c^i/\sigma^i$ is the $i-$region recharging time, and
 $c^i$ is the geometrical capacitance per unit width of the device   determined mainly 
 by the dielectric constant $\kappa$ of the isolating  material surrounding the GL, and geometrical parameters~(see, for example,~\cite{36,37,38,39}). If the GL is deeply embedded into the isolating material $c^i \propto \kappa$. In the case of free GL top surface between the gates, $c^i \propto \kappa^{eff} = 
 (\kappa +1)/2$. A thin passivation layer can also affect $c^i$.
    
According to Eq.~(1), 

\begin{eqnarray}\label{eq3}
 \sigma^i =  b\frac{e^2}{\hbar}\sqrt{\frac{e{\overline \Phi}}{2l\hbar\,v_W}},
\end{eqnarray}
where $b = 3a/2 \simeq 0.0716$. 
The quantity  
 
\begin{eqnarray}\label{eq4}
 F_{\omega} ={\mathcal J}_0 (\omega\,t^i/2)\, e^{i\omega\,t^i/2}
\end{eqnarray}
reflects the signal frequency dependence of the $i$ region dynamic conductivity determined by
the ac current induced by the carriers propagating between the $p$ and  $n$ regions (and the pertinent gates), particularly, by the finiteness of their transit time
$t^i =2l/v_W$, where 
 ${\mathcal J}_0(s)$ is the Bessel  function of the first kind. 

Using Eqs.~(2) and (4), we obtain the following expressions for the real and imaginary parts of $\sigma^i_{\omega}$:
 
\begin{eqnarray}\label{eq5}
 {\rm Re}~\sigma^i =  \sigma^i{\mathcal J}_0 (\omega\,t^i/2)\, \cos(\omega\,t^i/2),
 \end{eqnarray}
 
 \begin{eqnarray}\label{eq6}
 {\rm Im}~\sigma^i =  \sigma^i[{\mathcal J}_0 (\omega\,t^i/2)\, \sin(\omega\,t^i/2) - \omega \tau^i]. 
\end{eqnarray}
The quantity   Re~$\sigma^i$ can be both positive and negative. However, at $\omega\,t^i <  \pi$,
Re~$\sigma^i > 0$.  In the range $\omega\,t^i \lesssim 4.8$, where $\sin(\omega\,t^i/2)> 0 $,  the first term on the right-hand side of Eq.~(6) is positive. The latter implies
that this term corresponds to the kinetic inductance of the holes and electrons in the $i-$region. 
In certain ranges of elevated frequencies, the product  ${\mathcal J}_0 (\omega\,t^i/2) \sin(\omega\,t^i/2)$ can be positive, although both factors are negative.

 For  $2l = 0.2~\mu$m and ${\overline \Phi} = 100 - 200$~mV, Eqs.~(1) and (4) yield
${\overline J}^i \simeq 0.34 - 0.95$~A/cm and  $\sigma^i$  in the range from 520 -
to  733~S/m (from 4.67-6.59~ps$^{-1}$). In the case of a sufficiently short $i$ region, the ballistic transport
takes place even at higher voltages
 ${\overline \Phi} > \hbar\omega_0/e$ despite the spontaneous emission of optical phonons, and
 the current density through this region and, hence, the differential conductance can substantially exceed the above estimates.

Considering  Eq.~(2), we arrive at the following equation for the linear ac current density in the $i$ region
expressed via the ac potential drop $\delta \Phi_{\omega}=(\delta \varphi_{\omega}^i|_{x=l} - \delta \varphi_{\omega}^i|_{x=-l})$ across this region:

\begin{eqnarray}\label{eq7}
 \delta J_{\omega} = \sigma_{\omega}^i \delta \Phi_{\omega}
\end{eqnarray}

\subsection*{Plasmonic response  of the gated regions} 
 
To express  $(\delta \varphi_{\omega}^i|_{x=l} - \delta \varphi_{\omega}^i|_{x=-l})$ via $\delta V_{\omega}$, we find the spatial distributions of the ac potential in the gated region accounting for
its  nonuniformity associated with the plasmonic effects. 
 
For  the densities of the  ac current in 
  $p$ and $n $regions ($ -L -l < x < -l$ and $l < x l+ L$), we have 
  
\begin{eqnarray}\label{eq8} 
\delta J_{\omega}^{g} =  -\sigma_{\omega}^g\,L
\frac{d \delta\varphi_{\omega}}{dx}\biggr|_{x = \pm l},
\end{eqnarray}
where $\sigma_{\omega}^g = \sigma^g [i\nu/(\omega+ i\nu)]$ and 
$ \sigma^g = (e^2\mu/\pi\hbar^2\nu\,L)$ are the $p$ and $n$ regions (gated) ac and  dc conductances, respectively,  $\nu$ is the frequency of hole and electron  momentum relaxation on impurities, acoustic phonons, 
 and $\delta \varphi_{\omega} = \delta \varphi_{\omega}(x, y)|_{y=0}$ expresses the ac potential spatial  distribution  along the  $x$ axis directed in the GPIN-FET-channel plane ($y=0$).   The frequency dependence given by Eq.~(8) accounts for  the kinetic inductance  of the gated regions.

For the gated sections of the channel [see Fig.~1(a)], we solve the linearized hydrodynamic equations for the hole and electron  plasmas in the related $p$ and $n$ regions,  disregarding the nonuniformity of the dc potential and carrier density distributions, and  arrive at the following equation for the spatial distribution of the ac potential $\delta \varphi_{\omega}$ accounting for the plasmonic response of the gated regions (see~\cite{12} and the references therein):

\begin{eqnarray}\label{eq9}
\frac{d^2\delta \varphi_{\omega}}{dx^2} +\frac{\omega(\omega + i\nu)}{s^2}\biggl(\delta \varphi_{\omega} \mp
\frac{\eta}{2}\delta V_{\omega}\biggr)= 0. 
\end{eqnarray}
Here the upper/lower sign is related to the $n$ region/$p$ region,  $s = \sqrt{4e^2\mu\,w/\kappa\hbar^2}$ is the plasma velocity, and $\mu = \mu^p=\mu^n$ is
the carrier Fermi energy in the gated regions of both types.  

The quantity $\eta = C_{cg}/(C_{cg} + C_g)$ characterizes the contact-gate coupling, where
 $C_{cg} \propto \kappa$ [or $C_{cg} \propto \kappa^{eff} = (\kappa +1)/2$]
   and $C_g =(\kappa\,L/4\pi\,w)$ 
 being the pertinent capacitances for the bladeike contacts and gates, where $w$ is the gate layer thickness.
 Normally, $C_{cg} \ll C_g$, hence $\eta \ll 1$.

Equation~(9) governs the ac potential in the $p $region ($-L -l <x <-l$) and in the $n$ region ($l < x < l+L$). It  accounts for both the gate and contact-gate capacitances  and the kinetic inductance of the holes and electrons in the gated regions. 

The boundary conditions  at the edges of the gated regions are:
 
\begin{eqnarray}\label{eq10}
\delta \varphi_{\omega}|_{\pm (l+L)} = \pm\frac{1}{2}\delta V_{\omega},\,
\end{eqnarray}

\begin{eqnarray}\label{eq11}
\sigma_{\omega}^i(\delta \varphi_{\omega}|_{x=l} -\delta \varphi_{\omega}|_{x=-l})=
\sigma_{\omega}^g\,L
\frac{d \delta\varphi_{\omega}}{dx}\biggr|_{x = \pm l}.
\end{eqnarray}

\begin{table*}[t]
\caption{\label{table} GPIN-FET parameters (*                                                           samples with identical parameters) \\ }  
\vspace{2 mm}
\centerline{\begin{tabular}{lccccccccccc} 
\hline\hline
Sample\, & $2l$~($\mu$m)& $t^i$~(ps) & $\tau^i$~(ps)&$L$~($\mu$m)& $w$~(nm)& $\kappa$ &  $\mu$~(meV)&  $\nu$~(ps$^{-1}$)  
&$\sigma^g/2\sigma^i$ 
 &$\Omega/2\pi$~(THz) 
\\ \hline \hline
a-1*&0.2 & 0.2  & 0.1 &1.0& 15 & 4.5& 100 &1.0 & 12&1.0 \\
a-2&0.2 & 0.2  & 0.1 &1.0& 15 &4.5 & 100 &2.0 & 6 &1.0  \\
a-3&0.2 & 0.2  & 0.1 &1.0& 15 &4.5 & 100 &3.0 & 4 &1.0 \\
\hline
b-1*&0.2 & 0.2  & 0.1&1.0& 15  &4.5 & 100 &1.0 & 12 &1.0 \\
b-2&0.1 & 0.1  & 0.1&1.0& 15  &4.5 & 100 &1.0 & 8.5 &1.0  \\
b-3&0.4 & 0.4  & 0.1&1.0& 15 &4.5 & 100 &1.0 & 16.9 &1.0\\
\hline
c-1*&0.2 & 0.2  & 0.1 &1.0& 15 &4.5 & 100 &1.0 & 12 &1.0 \\
c-2&0.2 & 0.2  & 0.1 &1.0& 15 &4.5 & 144 &1.0 & 14.4 &1.2\\
c-3&0.2 & 0.2  & 0.1 &1.0& 15 &4.5 & 64 &1.0 & 9.6& 0.8\\
\hline
d-1*&0.2 & 0.2  & 0.1 &1.0& 15 &4.5 & 100 &1.0 & 12&1.0 \\
d-2&0.2 & 0.2  & 0.2 &1.0& 33 & 10.0  & 100 &1.0 & 12&1.0 \\
d-3&0.2 & 0.2  & 0.3 &1.0& 50 & 15.0 & 100 &1.0 & 12&1.0 
\\ 
\hline\hline
\end{tabular}}
\end{table*}

Using  the solution of  Eqs.~(9) - (11) [see Appendix B, Eqs.~(B11) and (B12)] and 
considering that $\delta \Phi_{\omega} = (\delta \varphi_{\omega}|_{x=l} -\delta \varphi_{\omega}|_{x=-l})$  we obtain

\begin{eqnarray}\label{eq12}
\delta \Phi_{\omega} = P_{\omega}\delta V_{\omega}, \qquad \delta J_{\omega } =  \sigma_{\omega}^i P_{\omega} \delta V_{\omega}
\end{eqnarray}
with $\delta V_{\omega} = \delta U_{\omega}$.
Here

\begin{eqnarray}\label{eq13}
P_{\omega} =  \frac{\eta\cos(\ae_{\omega}L) + 1 - \eta}
{\cos(\ae_{\omega}L)+ \displaystyle\frac{1}{\xi_{\omega}}\frac{\sin(\ae_{\omega}L)}{(\ae_{\omega}L)}}. 
\end{eqnarray}
Here the following notations have been introduced:
 $\ae_{\omega} =\pi\sqrt{\omega(\omega+i\nu)}/2\Omega\,L$, 
$\xi_{\omega} = \sigma_{\omega}^g/2\sigma_{\omega}^i = (\sigma^g/2\sigma^i)[i\nu/(\omega+i\nu)(F_{\omega}- i\omega\tau^i)]$. The quantity  

\begin{eqnarray}\label{eq14}
\Omega =  \frac{\pi\,e}{\hbar\,L} \sqrt{\frac{\mu\,w}{\kappa}}
\end{eqnarray}
is  the plasma frequency of the gated regions (see for example,~\cite{12,40,41,42,43}).

Equation~(14) accounts for the fact 
 that the electron liquid in the gated portion of the channel at the carrier densities under consideration is 
 degenerate.

Considering that in the real case $\eta \ll 1$, the function $P_{\omega}$ is weakly sensitive to the contact-gate capacitive coupling, in particular, to the parasitic
capacitance $C_{cg}$ (see Sec.~VII), and  Eq.~(13) can be somewhat simplified as

\begin{eqnarray}\label{eq15}
P_{\omega } \simeq  \frac{1 }
{\cos(\ae_{\omega}L)+ \displaystyle\frac{1}{\xi_{\omega}} \frac{\sin(\ae_{\omega}L)}{(\ae_{\omega}L)}}.
\end{eqnarray}.

\begin{figure}[t] \centering
\includegraphics[width=8.0cm]{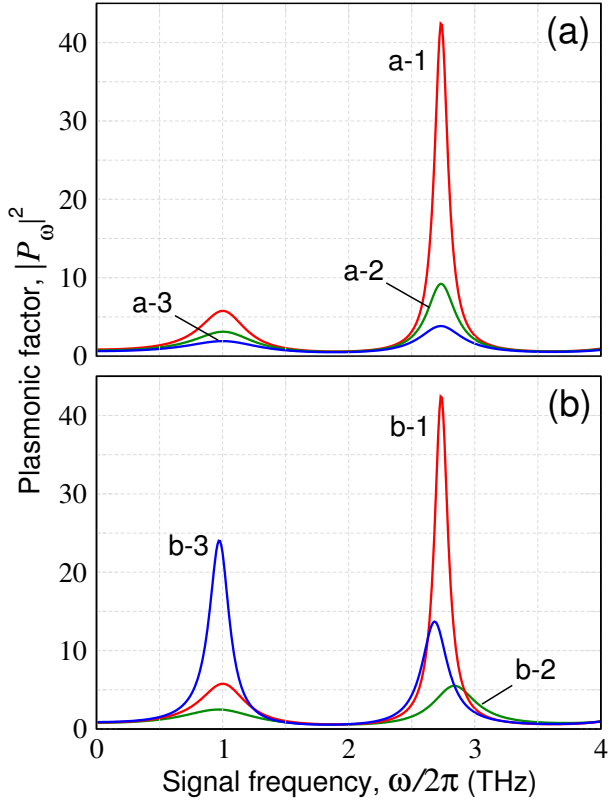}\\
\caption{Plasmonic factor $|P_{\omega}|^2$ versus signal frequency $f = \omega/2\pi$ 
 (a) for GPIN-FETs (a-1) - (a-3) with different values of frequency, $\nu$, of the carrier momentum relaxation in the gated regions and, consequently, different conductivity, $\sigma^g$, of these regions and (b) for  GPIN-FETs (b-1) - (b-3)  with different lengths of the $i$ region $2l$, corresponding to different differential conductivity  $\sigma^i$, and transit time  $t^i$~ 
 (${\overline U} = 200$~mV). The parameters of the samples  are given in Table I.    
} 
\label{F2}
\end{figure}

\section{Rectified current and plasmonic  factor}

The $I-V$ characteristic given by Eq.~(1) corresponds to 
 the nonlinear current density component, which comprises the rectified component and the ac current harmonics:

\begin{eqnarray}\label{eq16}
 \Delta {\tilde J}_{\omega}= \beta\, \delta\Phi_{\omega}^2,
\end{eqnarray} 
where 

\begin{eqnarray}\label{eq17}
 \beta = \frac{1}{2}\,\frac{d^2 {\overline J}}{d\, {\overline \Phi}^2} =  \frac{\sigma^i}{4{\overline \Phi}} = \frac{be^{5/2}}{4\sqrt{2l\hbar^{3}v_W {\overline \Phi}}} \propto \frac{1}{\sqrt{\overline \Phi}}
\end{eqnarray} 
is the parameter characterizing the nonlinearity of the $i-$region tunneling $I-V$ characteristics given by Eq.~(1), i.e., the  curvature of these characteristic.

According to the Kirchhoff law,

\begin{eqnarray}\label{eq18}
{\overline  \Phi}  =
\frac{{\overline U}}{\gamma},\qquad \delta V_{\omega} = \delta U_{\omega},
\end{eqnarray} 
where $\gamma =1  + 2H\sigma^ir^{load}/3$.
Here we accounted for the fact  that the $i$ region dc conductance is equal to $2\sigma^i/3$.

Equation~(16) yields the following expression for the density of the rectified current
$\Delta  {\overline J}_{\omega} = (\omega/2\pi)\int_{0}^{2\pi/\omega}d t \Delta {\tilde J}_{\omega}$:

\begin{eqnarray}\label{eq19}
\Delta {\overline J}_{\omega}  =
\frac{\sigma^i\gamma}{8{\overline U}}|P_{\omega}|^2\delta U_{\omega}^2.
\end{eqnarray}

As seen from Eq.~(19), the rectified current density $\Delta{\overline J}_{\omega}$, as will be seen
in the following, and the frequency dependence of the GPIN-FET detector responsivity, 
are determined by the plasmonic factor $|P_{\omega}|^2$. According to Eq.~(16), this factor is given by:

\begin{eqnarray}\label{eq20}
|P_{\omega}|^2 =  \biggl|\frac{\eta\cos(\ae_{\omega}L) + 1 - \eta}
{\cos(\ae_{\omega}L)+ \displaystyle\frac{1}{\xi_{\omega}}\frac{\sin(\ae_{\omega}L)}{(\ae_{\omega}L)}}\biggr|^2. 
\end{eqnarray}

Figure~2 shows the frequency dependences of the plasmonic factor $|P_{\omega}|^2$ calculated
for the GPIN-FETs with $L= 1.0~\mu$m, $w = 50$ nm, $\mu = 100$~meV, $\kappa = 4.5$, $\tau^i \simeq 0.1$~ps$^{-1}$($c^i \simeq 0.5$), $\Omega/2\pi = 1.0$~THz, and  $\eta = 0.1$ at  ${\overline \Phi} = 100$ mV, i.e., $\overline U \simeq 200$~mV and  $\gamma \simeq 2$ (see  Table I).

As seen in Fig.~2,  $|P_{\omega}|^2$ and, consequently, 
 the rectified current components given by Eqs.~(19), exhibit two 
 pronounced maxima (in the frequency range under consideration, $\omega/2\pi \leq 4$~THz). 
 
 The first maximum at $\omega/2\pi \simeq  1$~THz corresponds to the signal frequency 
 close to the gated regions plasma frequency $\Omega/2\pi$. It is obviously related to the excitation
 (by the incoming signal)
 of the fundamental mode of standing plasma wave  with the wave number $q_1 \simeq \pi/2L$ and
  the  maximum amplitude at $x=\pm l$  having the opposite phases.
 As a result, $|\delta \Phi_{\omega}|$ and $|\delta J_{\omega}|$ are maximal.
 The collisional damping of this plasma oscillations results in lowering of the plasma resonance peaks
 [see the curves corresponding to different $\nu$ in Fig.~2(a)].

The second resonant  peaks of $|P_{\omega}|^2$ at $\omega/2\pi \lesssim 3$~THz seen in Fig.~2 correspond to the excitation of the 
 plasmonic  mode with the wave number $q_3 \simeq 3\pi/2L $. 
It is instructive that the peaks associated with this mode  can be higher than those related to the fundamental mode. 

At elevated signal frequencies Eqs.~(19) and (20) yield

\begin{eqnarray}\label{eq21}
\Delta {\overline J}_{\omega} \propto |P_{\omega}|^2 \propto \biggl(\frac{{\overline \omega}}{\omega}\biggr)^2,
\end{eqnarray}
where 
$$
{\overline \omega} = \frac{1}{\tau^i}\biggl(\frac{\pi\nu}{2\Omega}\biggr)\biggl(\frac{\sigma^g}{2\sigma^i}\biggr). 
$$
For the parameters of samples (a-1) - (a-3), we obtain ${\overline \omega}/2\pi \simeq 4.8$~THz.
The roll-off of $\Delta {\overline J}_{\omega}$ and  $ |P_{\omega}|^2$ with increasing frequency
is determined by the charging time $\tau^i$, i.e., is associated with the $i-$region geometrical capacitance $c^i$. Considering that $\sigma^g \propto 1/\nu$, one can find that ${\overline \omega}$ is independent of $\nu$.

\section{GPIN-FET detector responsivity}

The current responsivity (ampere-watt)  of  GPIN-FETs operating as the terahertz detectors  is given by

\begin{eqnarray}\label{eq22}
R_{\omega}^J = \frac{\Delta {\overline J}_{\omega} H}{SI_{\omega}},
\end{eqnarray}
where $H$ is the lateral size of the GPIN-FET in the direction perpendicular to the gate edges,
$S = \lambda^2_{\omega}g/4\pi$ and $g \sim 1.5$ are the antenna aperture and gain,  $\lambda_{\omega} = 2\pi\,c/\omega$ and $I_{\omega}$ are the wavelength and intensity of the incoming radiation, and $c$ is the speed of light in vacuum. Considering that $\delta U_{\omega}^2 = 8\pi\lambda^2_{\omega}I_{\omega}
/c$, using Eq.~(22), we obtain

\begin{eqnarray}\label{eq23}
R_{\omega}^J = {\overline R}^J |P_{\omega}|^2, \qquad {\overline R}^J = \frac{4\pi^2H\sigma^i}{cg}{\frac{\gamma}{{\overline U}}}.
\end{eqnarray}

For the voltage (volt-watt) responsivity $R_{\omega}^V = r^{load}R_{\omega}^J$, 
assuming that the load resistance is optimized ($r^{load} =3/2\sigma^iH$, i.e., $\gamma \simeq 2$ if
$\sigma^i \ll \sigma^g/2$), we arrive at the following
universal formula:

\begin{eqnarray}\label{eq24}
R_{\omega}^V = {\overline R}^V | P_{\omega}|^2,\qquad {\overline R}^V = \frac{6\pi^2\gamma}{cg{\overline U}}.
\end{eqnarray}
For ${\overline U} = 200 - 400$~mV, we obtain ${\overline R}^V \simeq (1.12 -0.56)\times10^4$~V/W. 
If  $2l = 0.2~\mu$m, $H = 10~\mu$m, and ${\overline U} = 200 - 400$~mV, the optimized load resistance
is equal to $r^{load} \simeq 307 - 216~\Omega$.
Naturally, the plasmonic resonance  may lead  to $R_{\omega}^V \gg {\overline R}^V$.

A decrease in   ${\overline R}^V$ and,  hence, in  $R_{\omega}^V$
with the rise of the bias dc voltage ${\overline U}$ 
is attributed to a decrease in the current-voltage characteristic nonlinearity parameter $\beta$ [see Eq.~(17)], and to a decrease in the channel resistance with rising bias voltage $\overline U$. The latter requires the pertinent decrease in  
the optimized load resistance $r^{load}$. 
However, lowering of the bias voltage is limited by the thermionic and thermogeneration processes in the $i$ region.

Figure~3 shows the GPIN-FET responsivity $R_{\omega}^V$ as a function of the signal frequency
$\omega/2\pi$ calculated using Eqs.~(23) and (24) involving Eq.~(20) for the devices (c-1) - (c-3)
with the parameters presented in Table I assuming ${\overline U} = 200$~mV.

First of all, one can see that GPIN-FETs can reveal fairly high
peak responsivity. Second, the responsivity peaks corresponding to the  higher plasma modes can be markedly higher
than the fundamental peak (see the discussion in Sec.~VI). Third, the second peaks  (c-1 and c-2) are positioned at the frequencies  $\omega/2\pi$ somewhat  lower than $\Omega/2\pi = 1.0$~THz and 
$\Omega/2\pi = 1.2$~THz, respectively. As mentioned in Sec.~IV, this is due to the effect
of the $i$ region capacitance on the resonant frequency deviating it from the plasma frequency of the gated region $\Omega$. The same is valid for the second and third peaks c-3 in the main plot and the inset.

\begin{figure}[t] \centering
\includegraphics[width=8.0cm]{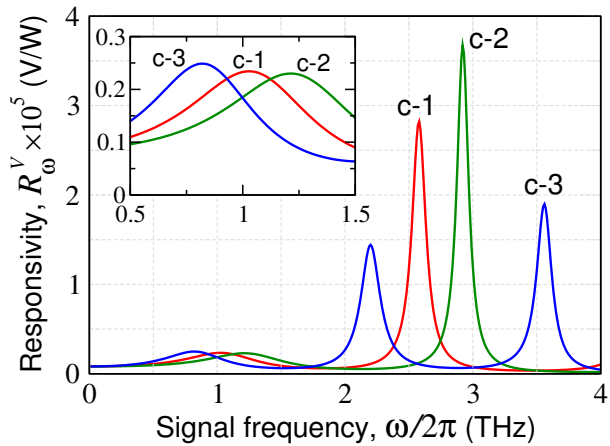}
\caption{ Frequency dependences of 
Volt-Watt responsivity  
for the GPIN-FETs (c-1) -(c-3) - with different values of the carrier Fermi energy $\mu$,
 conductance $\sigma^g$, and plasma frequency~$\Omega$ (${\overline U} = 200$~mV). Inset shows
 the vicinity of the fundamental resonance in more detail. 
} 
\label{F3}
\end{figure}

\section{ Discussion (Analysis)}
 
\begin{figure}[b] \centering
\includegraphics[width=8.0cm]{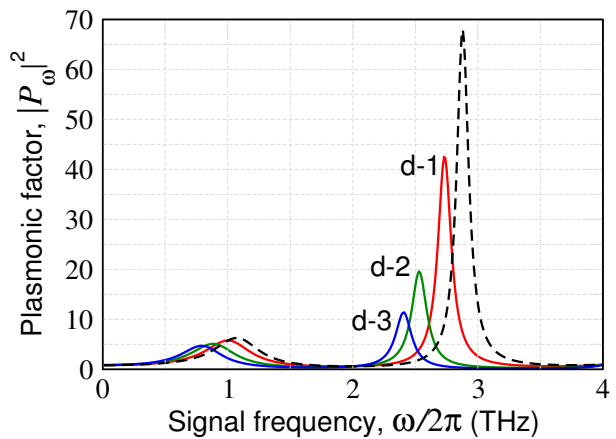}
\caption{ Plasmonic factor $|P_{\omega}|^2$ versus signal frequency $f =\omega/2\pi$  for GPIN-FETs (d-1) - (d-3), which differ by  charging time $\tau^i$ (proportional to  gate layer dielectric constants $\kappa$). The dashed line corresponds to   $\tau^i = 0.05$~ps ($\kappa = 2.25$).
} 
\label{F4}
\end{figure}

\subsection*{Temperature dependence and heights of the resonant peaks}

The obtained formulas for the GPIN-FET responsivity $R_{\omega}^V$ do not explicitly account for the temperature dependence,
at least at $\mu \gg T$, i.e., in the situations under consideration. 
This is because of the tunneling nature of the current in the GPIN-FETs.
However, the plasmonic resonant factor, which determines the maximum values of $R_{\omega}^V$ is sensitive to the collision frequency $\nu$. The latter is usually smaller at lower temperatures due to
a decrease in the carrier momentum relaxation on acoustic phonons. 
Hence lowering of the temperature can result in a marked sharpening of the resonant responsivity peaks and  promote a substantial increase in the resonant responsivity.

Figures~2 and 3, show some 
 deviation of the resonant  peaks position from the exact plasmonic resonances  $\omega/2\pi = \Omega/2\pi = 1$~THz and  $\omega/2\pi = 3$~THz -
 the peaks are shifted toward smaller frequencies.
This is  
 attributed to the collisional damping and to the contribution of the i-region geometrical capacitance.
Indeed, as seen from Fig.~4, the GPIN-FETs with longer charging time $\tau^i$, i.e., a larger $i$ region geometrical capacitance $c^i$ due to a larger dielectric constant of the gate layer $\kappa$ (and the same collisional frequency $\nu$)
exhibit smaller resonant frequencies.
The  gate layer in the samples (a-1) - (d-1) is assumed to be made of $h$-BN ($\kappa = 4.5$), whereas  the gate layers
in  the samples d-2 and d-3 are made of SiC ($\kappa = 10$) and HfO$_2$ ($\kappa = 15$), respectively.

Table I lists the values of $\kappa$ for the samples under consideration. 
The assumed value of $\kappa$ for GPIN-FET passivated by $h$-BN is close to its in-plane value because the in-plane direction
of the electric field .  The-high-frequency dielectric constant of crystalline $h$-BN in plane is $\kappa = 4.98$ and out-of plane $\kappa = 3.4$~\cite{44}. However, the exact effective value of $\kappa$ depends on the thickness of the top passivating layer.
GPIN-FETs could use passivating layers with very low values of $\kappa$, such as polyimide, porous BN/polyimide composites, or amorphous $h$-BN with the measured dielectric permittivity of 1.16  (close to that of air)~\cite{45,46,47}.
In Fig.~4,
we  also added the plot (see the dashed curve) corresponding to the parameters similar to those of the sample d-1, but with a shorter  charging time (smaller dielectric constant, $\kappa = 2.25$).
As seen, the pertinent peak is shifted weaker than others.   
This confirms that the $i-$region geometrical capacitance markedly affects the plasmonic resonances.

The most intriguing feature of the $|P_{\omega}|^2$ and $R_{\omega}^V$ frequency dependences is the larger height of the peaks corresponding to a higher plasmonic mode.  
Compare the plasmonic factors at the fundamental and second resonances.
The plasmonic factor $|P_{\omega}|^2$  near the fundamental ($\omega \simeq \Omega$) and the second 
($\omega \simeq 3\Omega$) resonances (at $\eta \simeq 0$), $\nu \ll \Omega$, and $3\Omega \tau^i\ll 1$
is equal to

\begin{eqnarray}\label{eq25}
|P_{\Omega}|^2 \simeq  \frac{(4\Omega/\pi\nu)^2}{\biggl|1 + \displaystyle\frac{8}{\pi}\biggl(\frac{2\sigma^i}{\sigma^g}\biggr){\mathcal J}_0(\Omega\,t^i/2)\,e^{i\Omega\,t^i/2}\biggr|^2}
\end{eqnarray} 
and

\begin{eqnarray}\label{eq26}
|P_{3\Omega}|^2 \simeq  \frac{(12\Omega/\pi\nu)^2}{\biggl|1- \displaystyle\frac{24}{\pi}\biggl(\frac{2\sigma^i}{\sigma^g}\biggr)\biggl(\frac{\Omega}{\nu}\biggr)^2{\mathcal J}_0(3\Omega\,t^i/2)e^{i\Omega\,t^i/2}\biggr|^2},
\end{eqnarray}
respectively. 

Considering that the product $\sigma^g \nu$ and, therefore, the denominators on the right-hand sides of Eqs.~(25) and (26) are independent of $\nu$, one can find that
$|P_{\Omega}|^2 \propto (\Omega/\nu)^2$ and $| P_{3\Omega}|^2\propto (\Omega/\nu)^2$. Such dependences
of the plasmonic factor peaks on $\nu$  are in line with the plots in Fig.~1(a).

As follows
from Eqs.~(25) and (26), the ratio of the peak heights corresponding to the fundamental and  second resonances,  at sufficiently large values $(2\sigma^i/\sigma^g)(\Omega/\nu)$, is equal to

\begin{eqnarray}\label{eq27}
n \simeq \biggl|\frac{P_{3\Omega}}{P_{\Omega}}\biggr|^2 \simeq \biggl[\frac{{\mathcal J}_0(\Omega\,t^i/2)}{{\mathcal J}_0(3\Omega\,t^i/2)}\biggr]^2.
\end{eqnarray}
This quantity can be  smaller or larger than unity, depending on $\Omega\,t^i$.
For all the samples considered above  (except b-3), $n > 1$; hence, the plasmonic resonances at $\omega\sim 3\Omega$ is stronger than those at $\omega \simeq \Omega$. 
This corresponds to
the curves (a-1) -(b-2) in Fig.~2. 
In contrast,  case b-3 is related to $n < 1$.
In particular, for samples a-1 and b-1, we obtain  $|P_{\Omega}|^2  \simeq 4.44$, $|P_{3\Omega}|^2 \simeq 42.37$, and $n \simeq 9.54$. This agrees well  with   the numerical calculation results
  shown in Fig.~2.

As mentioned above, 
the height of $|P_{\omega}|^2$ and $R_{\omega}^V$ resonant peaks   decreases with increasing collision frequency $\nu$. This is due to the strengthening  of the plasma oscillation collisional damping.
The hole and electron viscosity also   damp  the plasma oscillations, particularly,  those with  larger wavenumbers $q$. This might be a reason for some lowering of the resonant peaks corresponding to higher plasma oscillations modes. The effect of the viscosity can be accounted for by replacing $\nu$ by $\nu^{visc} = \nu+ hq^2$
where  $h$ is the electron viscosity ~\cite{1,32,48,49}. 
For the fundamental,   second, and third   plasma modes with $\omega_1 \simeq \Omega$, 
$\omega_2 \simeq 3\Omega$ and $\omega_3 \simeq 5\Omega$, the wavenumbers are 
$q_1 = \pi/2L$,  $q_2 = 3\pi/2L$, and $q_2 = 5\pi/2L$. 
If $L= 1~\mu$m, setting $\nu = 1 -2$~ps$^{-1}$
and $h = 100 -500$~cm$^2$/s
(depending on the carrier density and the temperature~\cite{32,48,49}),
$(\nu_1^{visc} - \nu)/\nu \simeq 0.0123 - 0.123$ and  $(\nu_2^{visc} - \nu)/\nu \simeq 0.11 - 1.11$, 
these estimates show that the viscosity effect can decrease the heights  of the second and fundamental 
peaks if the viscosity is sufficiently strong. As a result, the quantity $P_{3\Omega}$ (for example, for the samples a-1 and b-1) can vary from $|P_{3\Omega}|^2 \simeq 42$ at $h=0$ to $|P_{3\Omega}|^2 \simeq 34$  and $20$ at $h = 100$~cm$^2$/s and  $h = 500$~cm$^2$/s, respectively.
Lowering of the fundamental resonant peak is markedly smaller -- about (2 -12)~$~\%$.
However, the second resonance remains stronger even at  relatively high viscosity.

In addition to the carrier collisions and the plasma viscosity mentioned above, the damping
mechanisms include plasmon-plasmon interactions, radiative damping, scattering of plasmons on
defects, dopant-induced plasmon decay, and the interaction of the hole and electron plasma with
the carriers in highly conducting contacts (see, for example,~\cite{50,51,52,53,54,55,56,57,58}). The intrinsic lifetime of
plasmons in GLs with the carrier densities assumed in our calculations is about 20 - 120 ps ~\cite{51},
i.e., much longer than the characteristic collision time $\nu^{ -1}$ . In sufficiently short GLs, the plasmon
decay time due to the contacts is of the order of $\nu^{ -1}$~\cite{54}. These estimates justify the assumption
that in the GPIN-FETs under consideration the collisional and viscosity damping mechanisms
dominate.

As for the third resonant peaks with $\omega_3 \simeq 5\Omega$ and $q_3 \simeq 5\pi/2L$, 
the plasma oscillation damping due to the viscosity 
is  strong enough to lead to  the  peak extinction
 (the right most peak in Fig.~3) at $h = 500$~cm$^2$/s and higher.

\subsection*{Effect of the contacts and gates coupling}

As follows from Eq.~(13), the plasmonic factor depends to  some extent on the capacitive coupling
between the  side contacts and the gates.  Figure~4 shows examples of the $|P_{\omega}|^2$ versus signal frequency $\omega/2\pi$ for the GPIN-FETs with the parameters corresponding to the sample a-2
for $\eta = 0.1$ [as in Fig.~2(a)] and for $\eta = 0.2$ at ${\overline U} = 200$~mV. 
One can see that an increase in $\eta$ leads to resonant peaks  lowering. Although one needs to keep in mind that the value $\eta = 0.2$
corresponds to an overestimated contact-gate capacitance $C_{cg}$ compared to typical values for FETs.
For the comparison, we calculate  $|P_{\omega}|^2$ also for a similar device structure, but with the short-cut side contacts and gates. 
In the latter case,  one can    use the following formula [coinciding with Eq.~(20) with $\eta = 1$]:

\begin{eqnarray}\label{eq28}
|P_{\omega}|^2 =  \frac{1}
{\biggl|1+ \displaystyle\frac{1}{\xi_{\omega}}\frac{\tan(\ae_{\omega}L)}{(\ae_{\omega}L)}\biggr|^2}. 
\end{eqnarray}
The pertinent dependence is shown in Fig.~5 as a dotted line. As seen in the latter
case, the plasma resonance associated with the gated regions (at $\omega \sim \Omega$)
is suppressed, while the plasmonic  resonance at $\omega \sim 2\Omega$ is sufficiently pronounced although being relatively weak.

\begin{figure}[t] \centering
\includegraphics[width=8.0cm]{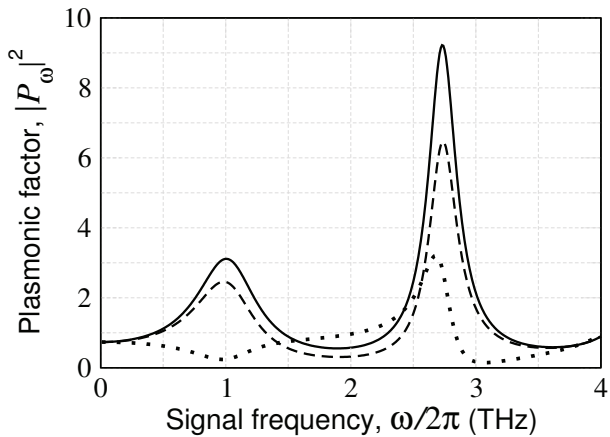}
\caption{ Plasmonic factor $|P_{\omega}|^2$ versus signal frequency $f =\omega/2\pi$ for the sample with the same parameters as the sample  a-2 and $\eta = 0.1$ (solid line) and $\eta = 0.2$ (dashed line). 
The dotted line is related to a GPIN-FET with the parameters a-2, but with the  short-cut side contacts and the pertinent gates.
} 
\label{F5}
\end{figure}

\subsection*{Thermionic and thermogeneration currents  in the reverse-biased $i$ region}

When $\overline \Phi \propto {\overline U}$ becomes relatively low,
the reverse thermionic current in the reverse-biased $i-$region can be comparable with the tunneling current.
Considering the tunneling current [given by Eq.~(1)] and the saturation current [given by Eq.~(C2)],            ],
 we find the following limitation for the minimal value of ${\overline U}$ at which the tunneling dominates over the thermionic processes:

\begin{eqnarray}\label{eq29}
e{\overline \Phi} > 
\biggl(\frac{6}{b\pi^2}\biggr)^{2/3} \biggl(\frac{2l\mu^2T^2}{\hbar\,v_W}\biggr)^{1/3}\exp\biggl(-\frac{2\mu}{3T}\biggr) = e {\overline \Phi}_{min}.
\end{eqnarray}
For $2l = 0.2~\mu$m, $\mu = 100$~meV, and $T = 25$~meV,
we obtain $e {\overline \Phi}_{min} \simeq 17.5$~meV.
This condition is satisfied in the above estimates and calculations.

The thermogeneration of the electron-hole pairs in the reverse-biased graphene $p-i-n$ junction is primarily associated with the interband absorption of optical phonons.
The thermionic rate at room temperature  is estimated as $g^{therm} = 10^{21}$~cm$^{-2}$s$^{-1}$~\cite{59}~. For $2l = 0.2~\mu$m, the latter yields
 $J^{therm} = 4elg^{therm} \simeq
6.4\times 10^{-3}$~A/cm.
For comparison, the tunneling current density in the reverse-biased $p-i-n$ junction given by Eq.~(1) for the same lengths of the $i$ region at $e{\overline \Phi} = 100 - 200$~meV, is equal to
${\overline J}^i \simeq 0.34  - 0.95$~A/cm, i.e., one order of magnitude larger.
This implies that the dark current in the GPIN-FETs in the conditions under consideration is determined by the interband tunneling in the $i$ regions.

\subsection*{Joule  heating}

The Joule power in the GPIN-FET is equal to 
$Q = H{\overline J}{\overline V}$. For ${{\overline \Phi} \lesssim \overline V} = 100 -200$~mV  (${\overline U} \simeq 200 - 400$~mV), 
${\overline J} \lesssim  0.5 - 1.0$~A/cm. Assuming the device width $H = 10~\mu$m, for
the  Joule power  we obtain $Q \simeq (5 - 10)\times 10^{-5}$~W. Since the bias voltage drops  primarily  in the i-region and somewhat around it, the Joule power releases in the area of about $2lH$.
This corresponds to the thermal power density $Q/2lH$. For $2JH = 2\times10^{-8}$~cm$^2$,
we obtain $Q/2lH \simeq 2.5 - 5.0$~kW/cm$^2$. The latter thermal power density is much lower than that, which can be  supported by GFETs (up to 210 kW/cm$^2$~\cite{60,61}). Hence, the Joule heating
should not lead to a marked overheating of the GPIN-FET channel.
Indeed,
the heat flow from the GPIN-FET channel through, for example, the $h$-BN substrate can be estimated considering the hBN thermal conductivity $k_{th} \simeq 20$~W/m$\cdot$K~\cite{62}. Assuming the thickness of the hBN layer $D = 1 -2~\mu$m, we for the thermal conductivity per unit area $K_{th} = k_{th}/D \simeq 
(1-2)$~kW/cm$^2$K. This implies that,  in such a case, the Joule heating results in   an increase 
of the channel temperature by $\Delta T \lesssim 5$~K.
Since graphene has a high room-temperature
thermal conductivity (about $k_{th} = 5$~kW/mK)~\cite{63},  the heat can also be effectively carried
to the side (metallic) contacts. 

\subsection*{Comparison of the Zener-Klein tunneling nonlinearity mechanisms with some other mechanisms }

It appears natural to compare the mechanisms of the current rectification in
 the GPIN-FETs under consideration and in   similar detectors (for example,~\cite{9}).
 In the latter devices, the 
 $I-V$ characteristic nonlinearity in the forward-biased graphene $p-n$
 junction is used. The pertinent nonlinearity parameter $\beta^{th}$ is given by Eq.~(C3).
 Comparing the quantities $\beta$ [see Eq.~(17)] and $\beta^{th}$ (for ${\overline \Phi} = - {\overline \Phi}^{th} < 0$ in the case of the detector using the $p-n$ nonlinearity) and $\overline {\Phi} >0$ for the GPIN-FET, we obtain

\begin{eqnarray}\label{eq30}
\frac{\beta}{\beta^{th}} 
\simeq \frac{\pi^2b}{8}\sqrt{\frac{\hbar\,v_W}{2l}}
\frac{T}{\mu}
\frac{\exp[(\mu-e{\overline \Phi}^{th})/T]}{\sqrt{e{\overline \Phi}}}
\end{eqnarray}
Setting $2l = 0.2~\mu$m, $\mu = 100$~meV, ${\overline \Phi} = 100$~mV, ${\overline \Phi}^{th} = 25$~mV,
  at $T = 25$~meV ($\simeq 300$~K )we obtain
  $\beta/\beta^{th} \simeq 0.1$. At the temperatures below room temperature, this ratio can be markedly larger. For example, at $T = (7.5 -10)$~meV [$\simeq (90-120)$~K], we obtain
 $\beta/\beta^{th} \simeq 1 - 9$.

\section{ Conclusions}
We demonstrate that the proposed GPIN-FETs detectors using the signal  rectification due to
the nonlinearity of the Zener-Klein tunneling $I-V$ characteristics can exhibit high responsivity
in the terahertz range of frequencies. The responsivity can be particularly high at the signal frequencies
close to the fundamental and triple carrier plasma frequency. This is due to the resonant excitation of plasma oscillation in the gate region of the GPIN-FET channel.
The GPIN-FETs can demonstrate competitive resonant responsivity at room temperatures. At lower temperatures, 
the responsivity can markedly increase due to the reinforcement of the resonant response because of
a weakening of the carrier momentum relaxation.

\section*{Acknowledgments}
The Japan Society for Promotion of Science (KAKENHI Grants No. 21H04546 and No. 20K20349), Japan; RIEC Nation-Wide Collaborative Research Project No. R04/A10; the US Office of Scientific Research Contract
N00001435, (project monitor Dr. Ken Goretta).

The authors declare no conflict of interest related to this article.

\section*{Appendix A} 
\setcounter{equation}{0}
\renewcommand{\theequation} {A\arabic{equation}}

\subsection*{Dynamic tunneling conductance of the reverse-biased  $i$ region}

The injected holes and electrons propagating in the $i$ region induce the current in the surrounding
highly conducting (gated) regions. The dynamical conductance, apart from the dc differential conductance 
$\sigma^i$ of the reverse-biased $i$region found from Eq.~(1),  should account for the displacement current   
associated with the geometrical capacitance $c^i$. Therefore, the  dynamical conductance of   the $i$ region associated with the propagating carriers
can be presented as

\begin{eqnarray}\label{eqA1}
\sigma_{\omega}^i = \sigma^i F_{\omega} -i\omega\,c^i.  
\end{eqnarray}
Because of a strong nonuniformity of the electric field in the $i-$region,
the tunneling generation of the electrons and holes occurs near the edges of the $p$ and $n$ regions, respectively. 
If the generated holes and electrons propagate across the $i$ region 
ballistically, their velocities are equal to $\pm v_W$.
Therefore, the ac electron and hole currents are proportional to
$\exp(i\omega\,x/v_W)$ and $\exp(-i\omega\,x/v_W)$. Hence, the ac current induced by the propagating  carriers is proportional
to 

$$
F_{\omega} = \frac{1}{2l}\int_{-l}^{l}dx g(x)[\exp(i\omega\,x/v_W)+ \exp(-i\omega\,x/v_W)],
$$
where, for the case of the device geometry under consideration (bladelike conducting areas), the form factor $g(x)$ is given by $g(x) = 2/\pi\sqrt{1 - (x/l)^2}$~\cite{64}.
This leads to

\begin{eqnarray}\label{eqA2}
F_{\omega} = \frac{2}{\pi}e^{i\omega\,t^i/2}
\int_{0}^{1}ds \frac{\cos(\omega\,t^is/2)}
{\sqrt{(1 - s^2)}}\nonumber\\
= e^{i\omega\,t^i/2} {\mathcal J_0}(\omega\,t^i/2),
\end{eqnarray} 
where $t^i =2l/v_W$ and ${\mathcal J}_0(s)$ is the Bessel function of the first kind.

If the potential drop $\overline \Phi > \hbar\omega_0$, the optical phonon emission can delay
the holes and electrons propagation in the $i$ region. For $\overline \Phi$ markedly exceeding 200~mV,  the spatial dependence
of the carrier velocity can be presented as $v_x = \pm v_W\exp(-x/\tau_{op}v_W)$, where $\tau_{op} \simeq 1- 2$~ps~\cite{58} is the time of the optical phonon spontaneous emission.  For the average transit time $<t^i> = (1/2l)\int_{-l}^ldx (2l/v_x)$, we obtain
$<t^i> = 2\tau_{op}\sinh(t^i/v_W\tau_{op}) \simeq t^i(1 + t^i/12\tau_{op}) \gtrsim t^i$.
For $2l = 0.2 - 0.5~\mu$m, the latter estimate yields $<t^i>/t^i -1 \gtrsim (1-4)\%$.

The $i$ region capacitance $c^i$, which determines the capacitive component of the displacement current through the $i-$region, 
for the bladelike conducting areas   can be estimated, generalizing ~\cite{36,65} by accounting for the presence of the highly conducting gates, as

\begin{eqnarray}\label{eqA3}
c^i =  \frac{\kappa}{2\pi^2} \Lambda,\qquad {\rm or}\qquad c^i =  \frac{\kappa^{eff}}{2\pi^2} \Lambda,
\end{eqnarray}
where $\Lambda$, as in ~\cite{36,52}, is a logarithmic factor, which   
 weekly depends on the geometrical parameters.
Equation~(A3)  
for the GPIN-FET  
with   $\kappa = 4.5$, yields  $c^i  \gtrsim 0.5$. This value is used in the main text.

\section*{Appendix B} 
\setcounter{equation}{0}
\renewcommand{\theequation} {B\arabic{equation}}

In the gradual channel approximation, the spatiotemporal distributions of the electron and hole densities, $\Sigma^n(x,t)$ and $\Sigma^p(x,t)$,
in the gated regions are related to the channel potential, $\varphi(x,t)$, and the gate potentials, $\varphi_g^n(t)$ and  $\varphi_g^p(t)$,
as

\begin{eqnarray}\label{eqB1}
\Sigma^n(x,t) = \frac{\kappa}{4\pi\,ew}[\varphi_g^n(t) - \varphi(x,t)],
\end{eqnarray}

\begin{eqnarray}\label{eqB2}
\Sigma^p(x,t) = \frac{\kappa}{4\pi\,ew}[\varphi_g^p(t)+ \varphi(x,t)],
\end{eqnarray}
where $\kappa$ and $w$ are the gate layer dielectric constant and thickness.
If the side source/drain contacts and the gates are coupled  by a capacitive link associated with
a free-space parasitic contact-gate capacitance [see Fig.~1(a)],

\begin{eqnarray}\label{eqB3}
\varphi_g^n(t) = V_g + \frac{C_{cg}}{C_g + C_{cg}}\frac{U(t)}{2},
\end{eqnarray}

\begin{eqnarray}\label{eqB4}
\varphi_g^n(t) = -V_g - \frac{C_{cg}}{C_g + C_{cg}}\frac{U(t)}{2},
\end{eqnarray}
where $C_{cg}$ and  $C_g$ are the contact-gate and gate-channel capacitances, respectively.

For the dc electron and hole densities, from  Eqs.~(B1) and (B2) we obtain

\begin{eqnarray}\label{eqB5}
{\overline\Sigma}^n = \frac{\kappa}{4\pi\,e}\biggl[V_g +\frac{\eta}{2}
{\overline U} - \overline{\varphi}(x)\biggr]
\nonumber\\
 \simeq \frac{\kappa}{4\pi\,e}\biggl(V_g +\frac{\eta}{2}
{\overline U} \biggr) = {\overline \Sigma},
\end{eqnarray}

\begin{eqnarray}\label{eqB6}
{\overline\Sigma}^p = \frac{\kappa}{4\pi\,e}\biggl[V_g +\frac{\eta}{2}
{\overline U} + \overline{\varphi}(x)\biggr]\nonumber\\
 \simeq \frac{\kappa}{4\pi\,e}\biggl(V_g +\frac{\eta}{2}
{\overline U} \biggr){\overline \Sigma}
\end{eqnarray}
with $\eta = C_{cg}/(C_g + C_{cg})$.
Since normally the dc bias voltage  ${\overline V} <  {\overline U} \ll V_g $, in Eqs.~(B5) and (B6)
we omit the spatially nonuniform  terms with ${\overline \varphi}(x)$. 
 
For the Fourier ac components, Eqs.~(B1) - (B4) yield

\begin{eqnarray}\label{eqB7}
\delta\Sigma_{\omega}^n(x) = \frac{\kappa}{4\pi\,ew}\biggl[\frac{\eta}{2}
\delta\,V_{\omega} - \delta\varphi_{\omega}(x)\biggr],
\end{eqnarray}

\begin{eqnarray}\label{eqB8}
\delta\Sigma_{\omega}^p(x) = \frac{\kappa}{4\pi\,ew}\biggl[-\frac{\eta}{2}
\delta\,V_{\omega} + \delta\varphi_{\omega}(x)\biggr].
\end{eqnarray}
Since the  load resistance is shunt by a large capacitance and the drop of the ac potential across the load resistor is insignificant, $\delta U_{\omega} = \delta V_{\omega}$.

As usual, considering the linearized hydrodynamic equations, expressing the electron and hole average velocities,
$\delta u^n_{\omega}$ and $\delta u^p_{\omega}$,
via the ac electric field $-d\delta \varphi_{\omega}(x)/dx$, and substituting $\delta u^n_{\omega}$ and $\delta u^p_{\omega}$ into the continuity equation, we obtain

\begin{eqnarray}\label{eqB9}
\frac{d^2\delta\varphi_{\omega}}{d\,x^2} 
- \frac{m\omega(\omega +i\nu)}{e{\overline \Sigma}}\delta \Sigma^n =0,
\end{eqnarray}

\begin{eqnarray}\label{eqB10}
\frac{d^2\delta\varphi_{\omega}}{d\,x^2} 
- \frac{m\omega(\omega +i\nu)}{e{\overline \Sigma}}\delta \Sigma^p =0,
\end{eqnarray}
where $m$ is the fictitious electron/hole mass in graphene. Expressing $m$ and ${\overline \Sigma}$
via the carrier Fermi energy, $\mu$,  in the gated region, we arrive at Eq.~(9) in the main text.

If the side contacts and the gates are shortened, one can  again  obtain Eq.~(16),
but with $\eta = 1$. 

Solving Eq.(9) with the boundary conditions given by Eqs.~(10) and (11), we obtain
 \begin{widetext}
\begin{eqnarray}\label{eqB11}
\delta \varphi_{\omega} =  \frac{1}{2}\delta V_{\omega} \biggl\{\eta+ (1-\eta)\cos[\ae_{\omega}(x-l-L)]
+ \frac{\eta +(1-\eta)[\cos(\ae_{\omega}L - \xi_{\omega}(\ae_{\omega}L)\sin(\ae_{\omega}L)]}{\sin(\ae_{\omega}L) +\xi_{\omega}(\ae_{\omega}L)\cos(\ae_{\omega}L)] }\sin[\ae_{\omega}(x-l -L)]\biggr\},\qquad
\end{eqnarray}

 \begin{eqnarray}\label{eqB12}
\delta \varphi_{\omega}=-\frac{1}{2}\delta V_{\omega} \biggl\{ \eta+ (1-\eta)\cos[\ae_{\omega}(x+l+L)]
- \frac{\eta +(1-\eta)[\cos(\ae_{\omega}L) - \xi_{\omega}(\ae_{\omega}L)\sin(\ae_{\omega}L)]}{\sin(\ae_{\omega}L) +\xi_{\omega}(\ae_{\omega}L)\cos(\ae_{\omega}L) }\sin[\ae_{\omega}(x+l +L)]\biggr\}\qquad
 \end{eqnarray}
for the $n$ region  and $p$ region ), respectively.
%
Here $\ae_{\omega} =\pi\sqrt{\omega(\omega+i\nu)}/2\Omega\,L$, 
$\xi_{\omega} = \sigma_{\omega}^g/2\sigma_{\omega}^i = (\sigma^g/2\sigma_{\omega}^i)[i\nu/(\omega+i\nu)]= (\sigma^g/2\sigma^i)[i\nu/(\omega+i\nu)(F_{\omega}- i\omega\tau^i)]$, and

\begin{eqnarray}\label{eqB13}
\Omega =  \frac{\pi\,e}{\hbar\,L} \sqrt{\frac{\mu\,w}{\kappa}}
\end{eqnarray}
is the plasma frequency of the gated regions.
   
\end{widetext}

\section*{Appendix C} 
\setcounter{equation}{0}
\renewcommand{\theequation} {C\arabic{equation}}

The thermionic dc current density,
including both the hole and electron components, in the $i-$region of the GPIN-FETs under consideration  can be presented as

\begin{eqnarray}\label{eqC1}
{\overline J}^{th}= J_s\exp\biggl(-\frac{\mu}{T}\biggr)\biggl[\exp\biggl(-\frac{e{\overline \Phi}^{th}}{T}\biggr) - 1\biggr], 
\end{eqnarray}
where
\begin{eqnarray}\label{eqC2}
J_s  \simeq \frac{4eT\mu}{\pi^2\hbar^2v_W}\exp\bigg(-\frac{\mu}{T}\biggr)
\end{eqnarray}
is the graphene  $p-i-n$ junction saturation current density.  In Eq.~(C1), as in the main text, 
the $p-i-n$ junction reverse and forward  bias voltages ${\overline \Phi} > 0$ and  ${\overline \Phi}^{th} < 0$, respectively.

Equations (C1) and (C2) yield the following value of the $I-V$ characteristic nonlinearity parameter:

\begin{eqnarray}\label{eqC3}
\beta^{therm} = \frac{1}{2} \frac{d^2{\overline J}^{therm}}{d{\overline \Phi}^2}
= \frac{\sigma^{i,th}}{2T}\nonumber\\
 \simeq \frac{2e^3}{\pi^2\hbar^2v_W}\biggl(\frac{\mu}{T}\biggr)\exp\bigg(-\frac{\mu+e{\overline \Phi} }{T}\biggr).
\end{eqnarray}
Here $\sigma^{i,th}$ is the differential conductance.

\end{document}